\documentclass[12pt]{iopart}

\usepackage{graphicx}
\usepackage{textcomp}
\usepackage{color}
\usepackage{dcolumn}
\usepackage{bm}
\usepackage{amssymb}

\begin{document}

\title{Fully guided-wave photon pair source for quantum applications}

\author{P Vergyris, F Kaiser, E Gouzien, G Sauder, T Lunghi, S Tanzilli}
\address{Universit\'e C\^ote d'Azur, CNRS, Institut de Physique de Nice (InPhyNi), France}
\ead{sebastien.tanzilli@unice.fr}

\begin{abstract}
We report a fully guided-wave source of polarisation entangled photons based on a periodically poled lithium niobate waveguide mounted in a Sagnac interferometer.
We demonstrate the source's quality by converting polarisation entanglement to postselection-free energy-time entanglement for which we obtain a near-optimal $S$-parameter of $2.75 \pm 0.02$, \textit{i.e.} a violation of the Bell inequality by more than 35 standard deviations.
The exclusive use of guided-wave components makes our source compact and stable which is a prerequisite for increasingly complex quantum applications. Additionally, our source offers a great versatility in terms of photon pair emission spectrum and generated quantum state, making it suitable for a broad range of quantum applications such as cryptography and metrology.
In this sense, we show how to use our source for chromatic dispersion measurements in optical fibres which opens new avenues in the field of quantum metrology.
\end{abstract}

\maketitle
\tableofcontents

\section{Introduction}

Sources of photonic entanglement have led to remarkable advances in quantum information science~\cite{Tittel_Photonic_2001}.
In quantum key distribution, entangled photon pairs can be used for provably secure communication purposes~\cite{Ekert_QKD_1991,Jennewein_QKD_2000,Gisin_QKD_2002,Aktas_DWDM_2016}. Linear optical quantum processors may solve some computational tasks in significantly more efficient ways compared to classical counterparts~\cite{Steane_qcomp_1998,KLM_2001}.
Also, optical quantum metrology exploits correlations between two, or more, photons in order to perform phase-sensitive measurements with sub-shot-noise precision~\cite{Giovannetti_qmet_2011}.

In all the above-mentioned fields, experimental complexity is increasing rapidly which sets high demands on stability, efficiency and compactness of the employed quantum photonic sources.
In this framework, integrated and guided-wave approaches have demonstrated their huge potential since photons are tightly confined and well isolated from environmental influences~\cite{Tanzilli_Genesis_2012,Alibart_PPLN_2016}.

In principle, any photonic entanglement observable can be chosen for applications.
For quantum key distribution, energy-time entanglement has proven to be advantageous thanks to its inherrent robustness against environmental influences. On the other hand, the polarisation observable is the most natural one, as its detection necessitates only standard optics such as wave-plates and polarising beam-splitters (PBS). Despite its sensitivity to environmentally-induced polarisation drifts, it has been shown that polarisation entanglement can even be distributed over long distances in fibre networks thanks to active polarisation stabilisation schemes~\cite{Chen_polar_stab_2007,Xavier_polar_stab_2008}.
Ultimately, a photon pair source should offer the dynamical choice between either observable, in view of adapting the setup to the particular needs of the experiment.

Here, we demonstrate the realisation of such a flexible source. Our setup naturally generates paired photons entangled in energy-time and polarisation and we exploit advantageously the latter observable to reveal energy-time entanglement using a postselection-free method~\cite{Strekalov_postselection_free_1996}, as opposed to the standard method~\cite{Brendel_E_T_1992,Tanzilli_E_t_2002}. The potential of our approach for applications is further underlined by performing quantum-enhanced chromatic dispersion measurements in metre sized optical fibres.

\subsection{State of the art guided-wave polarisation entangled photon pair sources}

There is a large variety of polarisation entanglement sources taking advantage of guided-wave photonics.
Polarisation entangled photon pairs can be generated through four-wave mixing in silicon ring resonators~\cite{Suo_hyper_Si_ring_2015} or birefringent fibres~\cite{Li_SFWM_2005,Fulconis_SFWM_2007,MeyerScott_cross_spliced_2013}.
Here, the corresponding emission bandwidth is usually relatively narrow which compromises the source versatility. Additionally, in order to separate the pairs from the pump laser (and the related Raman scattering noise), several filtering stages are required which increases optical losses significantly, while the paired photons need to be at non-degenerate wavelengths~\cite{Fulconis_SFWM_2007,MeyerScott_cross_spliced_2013}.
On the other hand, guided-wave polarisation entangled photon pair sources based on spontaneous parametric downconversion (SPDC) offer noise-free and high-efficiency generation~\cite{Suhara_review_2009,Alibart_PPLN_2016}.
Exploiting type-II SPDC in periodically poled waveguides, such as lithium niobate (PPLN/W) and potassium titanyl phosphate, allows to naturally generate polarisation entangled photon pairs, however with a relatively pair narrow emission bandwidth~\cite{Suhara_type_II_2007,Martin_type_II_2010}.
Broader emission spectra offer increased wavelength flexibility, and, for example, the possibility to simultaneously provide many standard telecommunication wavelength channels with entangled photon pairs using only one source~\cite{Aktas_DWDM_2016,Lim2_Sagnac_2008,Autebert_DWDM_2016,Arahira_DWDM_2016}.
This can be obtained through type-0 SPDC in PPLN/W, via which, however, only the energy-time entanglement observable is naturally available~\cite{Sanaka_E_t_2001,Tanzilli_E_t_2002}. Therefore, generation of polarisation entanglement further requires quantum state engineering, \textit{e.g.} interferometric setups like a birefringent delay line~\cite{Takesue_PDL_2005,Kaiser_source2} or by having two type-0 generators in a Mach-Zehnder type configuration~\cite{Yoshizawa_MZI_2004,Herbauts_MZI_2013}.
In order to get rid of active interferometer stabilisation systems, several groups employed a PPLN/W in a Sagnac loop configuration~\cite{Odate_Sagnac_2007,Lim3_Sagnac_2010}. However, the related setups required some special free-space optics, compromising both stability and compactness.
Fully guided-wave approaches were demonstrated using a configuration in which second-harmonic generation and SPDC occur in the same Sagnac loop~\cite{Jiang_Sagnac_SHG_SPDC_2007,Arahira_Sagnac_SHG_SPDC_2011}. However, similarly to four-wave mixing approaches, only non-degenerate photon pairs can be produced and significant efforts for pump laser filtering are required.

In the following, we show the realisation of a photon pair source that overcomes the above-mentioned issues, \textit{i.e.} a source of broadband polarisation entangled photon pairs, entirely based on guided-wave photonics.
We note also that this source has been used within the framework of \textit{The Big Bell Test}~\cite{TBBT_webpage} and details on this particular experiment will be reported elsewhere.
Additionally, we demonstrate how to exploit the versatile design of our approach for performing chromatic dispersion measurements in optical fibres.

\section{Experimental setup}

\begin{figure}
\begin{center}
\includegraphics[width=1\columnwidth]{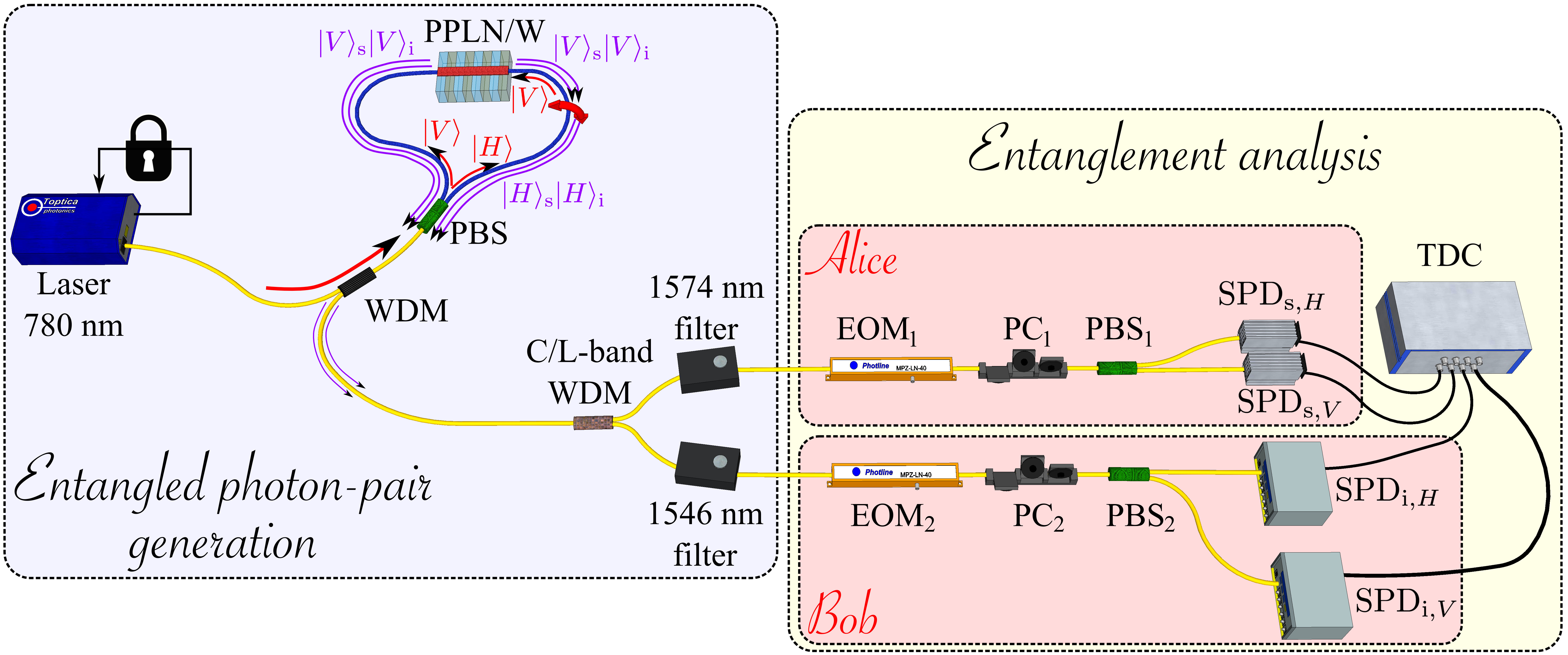}
\caption{Experimental setup. Energy-time and polarisation entangled photon pairs are created in a fully-guided wave configuration based on a Sagnac loop. The photon pair generator is a 3.8\,cm long PPLN/W. Standard single-mode fibres are symbolised by yellow lines while PMFs are represented by blue lines. The length of the right-hand side PMF can be adapted by inserting patch cords of different lengths.
Paired photons are generated in the maximally entangled state $| \psi_1 \rangle = \frac{1}{\sqrt{2}} \left( |V \rangle_{\rm s} |V \rangle_{\rm i} + |H \rangle_{\rm s} |H \rangle_{\rm i}\right)$. Signal and idler photons are deterministically separated at the C/L-band WDM and futher filtered down to a bandwidth of about 1\,nm. Two electro-optic phase modulators (EOM$_{1,2}$) are used to induce phase-shifts between the $|V \rangle_{\rm s} |V \rangle_{\rm i}$ and $|H \rangle_{\rm s} |H \rangle_{\rm i}$ contributions. Photons are projected into the diagonal basis using PC$_{1,2}$, PBS$_{1,2}$, four single photon detectors, and a TDC. As a result, energy-time entanglement is analysed in a postselection-free fashion~\cite{Strekalov_postselection_free_1996}. \label{Fig1}}
\end{center}
\end{figure}

The schematics of the experimental setup is shown in \figurename~\ref{Fig1}.
It is divided in two parts, \textit{i.e.} a photon pair generation stage, and an entanglement analysis stage.

\subsection{Entangled photon pair generation}

An actively wavelength-stabilised fibre-coupled 780\,nm continuous-wave pump laser is sent through a 780/1550 wavelength division multiplexer (WDM) to a fibre polarising beam-splitter (PBS, Photonik Singapore Ptd Ltd), defining the in- and output ports of the Sagnac loop. Although the PBS is optimised for the telecom wavelength band, its performance at 780\,nm is reasonably good, \textit{i.e.} the overall transmission is $\sim$50\% and the polarisation extinction ratio is $\gtrsim$30:1.
At the output of the PBS, vertically polarised pump laser photons propagate in the clockwise direction $(|V \rangle_{\rm p,\circlearrowright})$, and horizontally polarised ones in the counter-clockwise direction $(|H \rangle_{\rm p,\circlearrowleft})$ in two telecom polarisation maintaining fibres (PMF). Note that, after the PBS, photons propagate along the slow PMF axis in both output fibres.
The latter PMF is twisted by $90^{\circ}$, therefore pump photons in this fibre are also vertically polarised $(|H \rangle_{\rm p,\circlearrowleft}\rightarrow |V \rangle_{\rm p,\circlearrowleft})$.
By connecting different fibre patch cords, we can choose the length difference between the two PMFs from zero to 32\,m which will be of interest for chromatic dispersion measurements.
However, if not explicitly stated, both PMFs have the same optical length within $\pm$1\,cm and are connected directly to the photon pair generator, a 3.8\,cm long PPLN/W.
In this crystal, vertically polarised pump photons can be converted to vertically polarised photon pairs through type-0 SPDC, $|V \rangle_{\rm p} \rightarrow |V \rangle_{\rm s} |V \rangle_{\rm i}$. Here, the subscripts s and i denote signal and idler photons, respectively.

\begin{figure}
\begin{center}
\includegraphics[width=0.65\columnwidth]{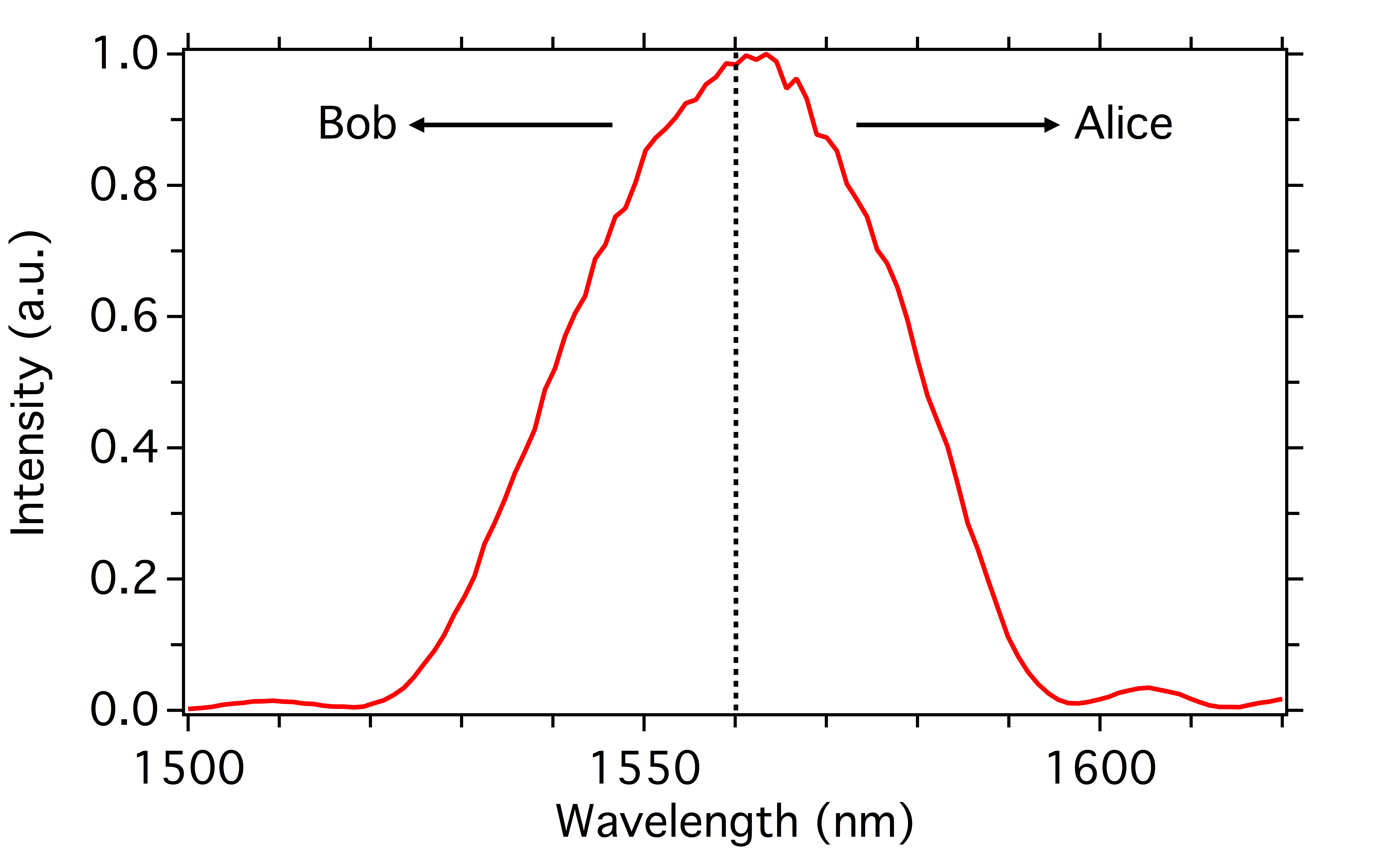}
\caption{Emission spectrum of the photon pair generator at a temperature of $26.0^{\circ}$C. The spectrum covers about 40\,nm at the degenerate wavelength. Using a C/L-band WDM, the spectrum is cut into two bands (vertical dashed line) in order to provide Alice (Bob) with the long (short) wavelength band.\label{Fig2}}
\end{center}
\end{figure}

By adjusting the polarisation of the pump laser photons, we can tune the probability of generating photon pairs in either direction.
The generated pairs are collected by the PMFs and further sent to the PBS. Note that the polarisation of the pair contribution generated in the clockwise direction is rotated by $90^{\circ}$ in the twisted PMF, \textit{i.e.} $|V \rangle_{\rm s,\circlearrowright} |V \rangle_{\rm i,\circlearrowright} \rightarrow |H \rangle_{\rm s,\circlearrowright} |H \rangle_{\rm i,\circlearrowright}$.
After the PBS, photon pair contributions generated in clockwise and counter-clockwise directions are recombined into the same spatial mode, and separated from the pump laser at the WDM, resulting in the formation of the following state
\begin{equation}
| \psi_1 \rangle = \alpha |V \rangle_{\rm s} |V \rangle_{\rm i} + {\rm e}^{i\,\phi}\,\beta |H \rangle_{\rm s} |H \rangle_{\rm i}, \qquad \rm with\,\,|\alpha|^2+|\beta|^2=1.
\end{equation}
Here, $\alpha$ and $\beta$ denote the probability amplitudes which are directly related to the pump laser polarisation.
The (constant) phase factor ${\rm e}^{i\,\phi}$ depends on the polarisation of the pump laser, the length of the PMFs, and, to a small extent, on environmental influences such as stress and temperature fluctuations in the PMFs.
By adjusting the pump laser polarisation properly, we can set $\alpha = \beta = \frac{1}{\sqrt{2}}$ and $\phi=0$, such that a maximally entangled Bell state is obtained:
\begin{equation}
| \Phi^+ \rangle =\frac{1}{\sqrt{2}} \left( |V \rangle_{\rm s} |V \rangle_{\rm i} + |H \rangle_{\rm s} |H \rangle_{\rm i} \right). \label{MaxEntState}
\end{equation} 
The wavelength-degenerate photon pair emission spectrum of our PPLN/W after the WDM is shown in \figurename~\ref{Fig2}. It covers about 40\,nm and both the shape and the central wavelength can be further tuned over a large range using basic temperature control~\cite{Kaiser_source2}.
As the photon pair generation process needs to obey the conservation of the energy, signal and idler photons are generated pairwise symmetrically around the degenerate wavelength of 1560\,nm.
For this reason, we choose to split the paired photons deterministically by sending the long (signal) and short (idler) wavelength parts to Alice and Bob, respectively.
Experimentally, this is implemented using a C/L-band WDM. Additionally, we employ 1\,nm tunable bandpass filters centred at 1574\,nm and 1546\,nm, respectively.
Note that our source is also capable of generating, in a non-deterministic fashion, degenerate entangled photon pairs which could be sent to Alice and Bob by replacing the WDM with a standard 50/50 beam-splitter, and adapting the bandpass filters accordingly.

\subsection{Entanglement analysis setup}

As the photon pairs are generated via continuous-wave SPDC, they are inherently energy-time correlated.
Usually this type of quantum correlation is analysed using a pair of unbalanced interferometers, referred to as the Franson configuration, giving  signal and idler photons a short and long path to the detector~\cite{Franson}.
Due to the spontaneous character of the photon pair generation process, the pair creation time remains unknown, making the two contributions in which both photons take the same path indistinguishable (short-short and long-long). 
Consequently, quantum interference may be observed as a function of the phase relation between these two contributions. This type of entanglement analysis necessitates, however, a postselection procedure in order to exclude contributions in which the paired photons take opposite paths (short-long and long-short)~\cite{Brendel_E_T_1992}, otherwise interference fringe visibilities are limited to 50\%~\cite{Ou_Bad_E_T_1990}.

Here, we take advantage of a previously reported strategy, developed by Strekalov \textit{et al.} towards analysing energy-time entanglement in a postselection-free fashion~\cite{Strekalov_postselection_free_1996}.
The basic idea is that the short-short and long-long contributions are replaced by $|V \rangle_{\rm s} |V \rangle_{\rm i}$ and $|H \rangle_{\rm s} |H \rangle_{\rm i}$, respectively.
As our particular photon pair source generates neither $|H \rangle_{\rm s} |V \rangle_{\rm i}$ nor $|V \rangle_{\rm s} |H \rangle_{\rm i}$ contributions, a postselection procedure therefore becomes unnecessary.

The phase relation between the two created contributions is tuned using fibre electro-optic phase modulators (EOM$_{1,2}$) with a half-wave voltage of about 2.5\,V. Depending on the voltage applied to the EOMs, a phase shift is applied to the $|H \rangle_{\rm s}$ and $|H \rangle_{\rm i}$ contributions, however, $|V \rangle_{\rm s}$ and $|V \rangle_{\rm i}$ remain always unaltered. Therefore, we obtain
\begin{equation}
| \psi_2 \rangle =  \alpha \, |V \rangle_{\rm s} |V \rangle_{\rm i} +e^{i\,\left( \phi_{\rm s} + \phi_{\rm i} \right)} \,\beta\, |H \rangle_{\rm s} |H \rangle_{\rm i}.
\end{equation}
Here, $\phi_{\rm s}$ and $\phi_{\rm i}$ are the phase shifts applied to the signal and idler photon contribution, respectively.

In order to observe interference with the highest visibility, the contributions to the photon pair state need to be indistinguishable. In our case, this is done in two steps. First, the probability amplitudes of those contributions are equalised by optimising the pump laser polarisation, \textit{i.e.} $| \alpha | = |\beta | = 1/\sqrt{2}$. Second, polarisation state information is erased by measuring the individual states (signal and idler) in the diagonal polarisation basis. This is done by rotating the polarisations of the corresponding states by $45^{\circ}$ using fibre polarisation controllers (PC$_{1,2}$), followed by quantum state projection on PBS$_{1,2}$ and four single photon detectors (SPD$_{{\rm s},V}$, SPD$_{{\rm s},H}$, SPD$_{{\rm i},V}$, SPD$_{{\rm i},H}$).
At Alice's site, we use free-running low-noise InGaAs SPDs (IDQ 220, 15\% detection efficiency, 20\,$\mu$s deadtime) which trigger, upon a detection event, Bob's gated InGaAs SPDs (IDQ 201, 20\% detection efficiency, 40\,$\mu$s deadtime).
All detectors are connected to a time-to-digital converter (TDC) which allows to infer the two-photon coincidence rates between different pairs of detectors.

\section{Results}

\subsection{Two-photon coincidence histograms}

In our arrangement, we expect two-photon coincidences between four different detector combinations, \textit{i.e.} SPD$_{{\rm s},V}$ \& SPD$_{{\rm i},V}$; SPD$_{{\rm s},V}$ \& SPD$_{{\rm i},H}$; SPD$_{{\rm s},H}$ \& SPD$_{{\rm i},V}$; and SPD$_{{\rm s},H}$ \& SPD$_{{\rm i},H}$.
\begin{figure}
\begin{center}
\includegraphics[width=0.75\columnwidth]{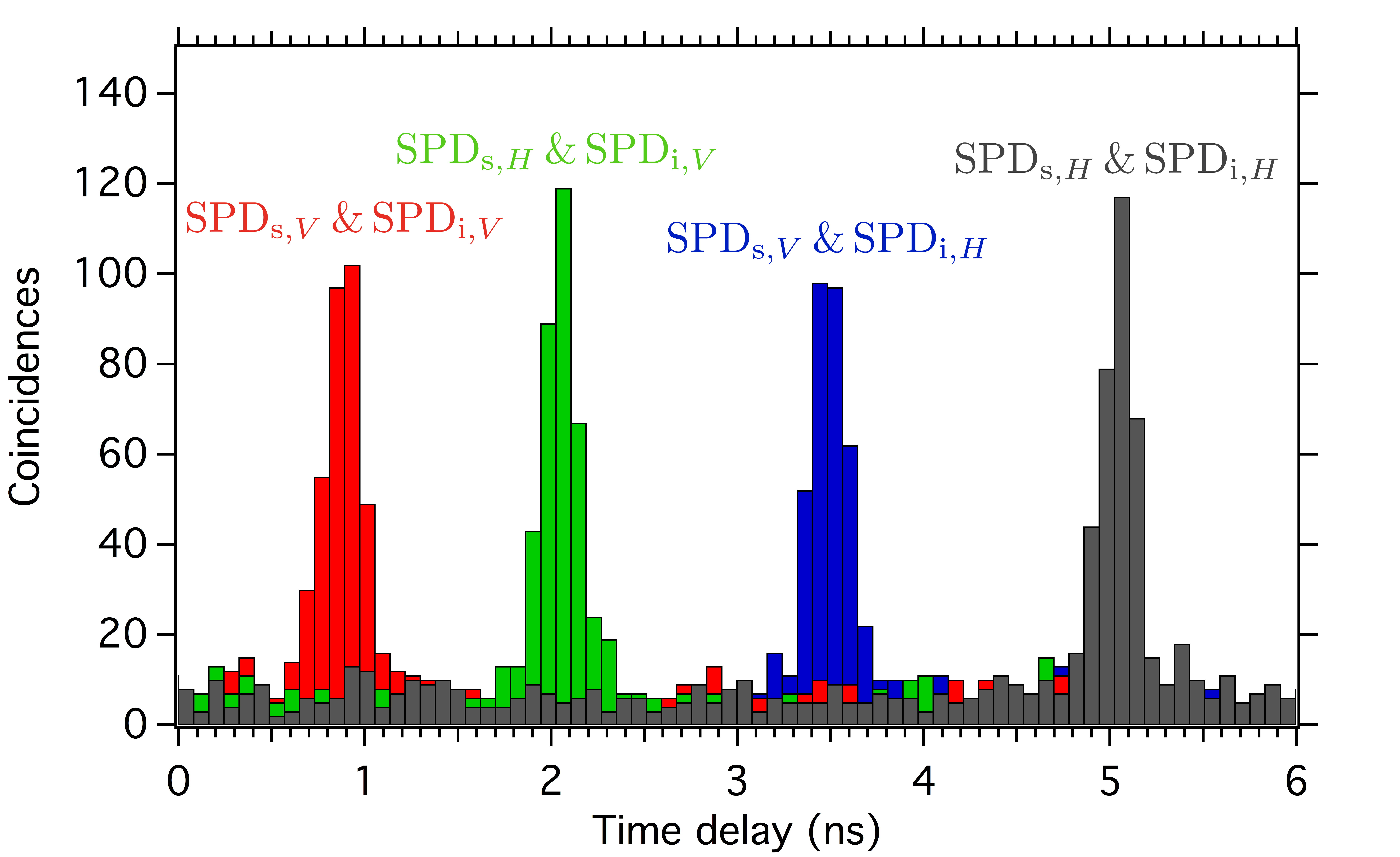}
\caption{Coincidence histograms for the four possible detector combinations, \textit{i.e.} SPD$_{{\rm s},V}$ \& SPD$_{{\rm i},V}$ (red); SPD$_{{\rm s},V}$ \& SPD$_{{\rm i},H}$ (blue); SPD$_{{\rm s},H}$ \& SPD$_{{\rm i},V}$ (green); and SPD$_{{\rm s},H}$ \& SPD$_{{\rm i},H}$ (grey). The temporal bin width of our TDC is 81\,ps. Note that, for better visualisation, the four peaks have been temporally displaced using different cable lengths.
Throughout the rest of the manuscript, only events within four bins located around the center of each coincidence peak are considered.\label{Fig3}}
\end{center}
\end{figure} 
Typical coincidence histograms, for random phase fluctuations applied to both EOMs, are shown in \figurename~\ref{Fig3}.
Four coincidence peaks are obtained, corresponding to the above-mentioned detector combinations.
To maintain an acceptable signal-to-noise level, only events within a 324\,ps region around the centre of each coincidence peak are considered.

\subsection{Entanglement analysis}

\begin{figure}
\begin{center}
\includegraphics[width=1\columnwidth]{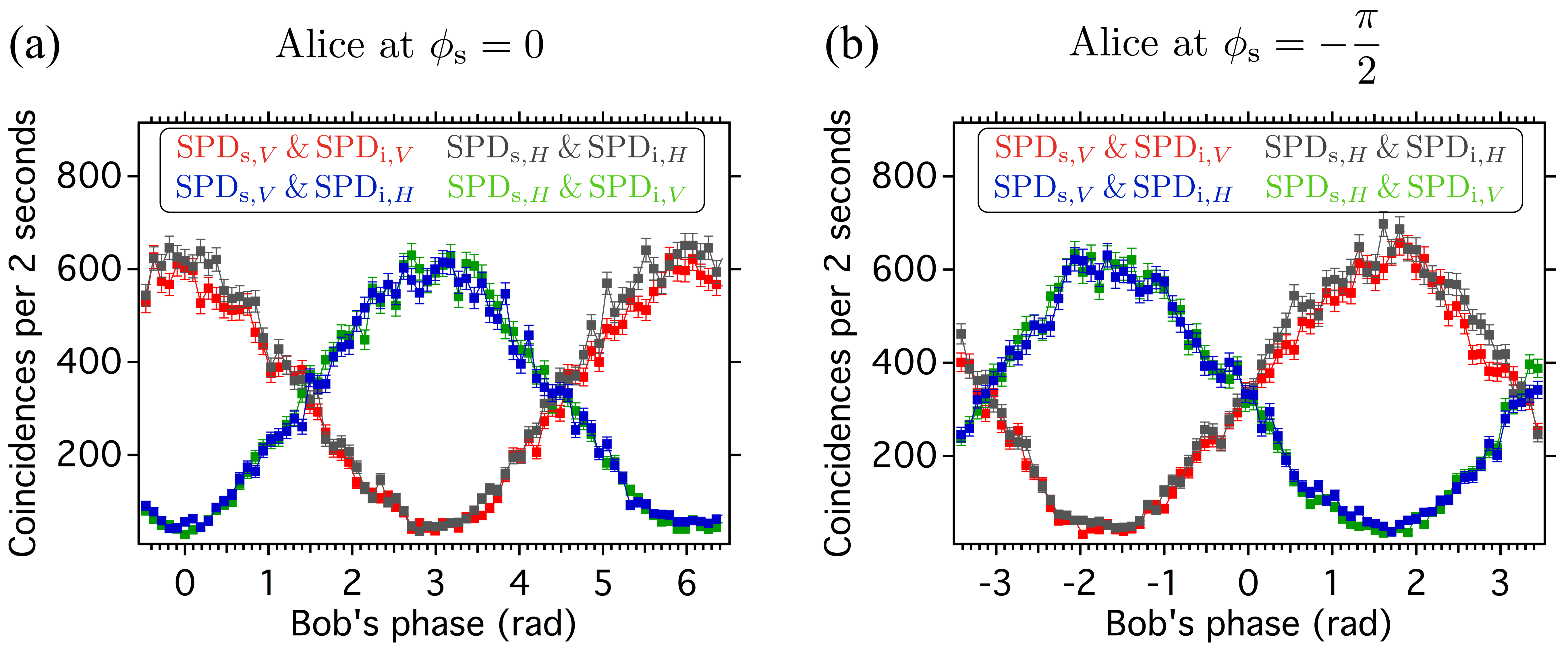}
\caption{Phase scans for the violation of the Bell inquality through postselection-free energy-time entanglement. At Alice's site, two fixed phase settings are used: (a) $\phi_{\rm s} = 0$ and (b): $\phi_{\rm s}= - \frac{\pi}{2}$. Bob's phase is continuously scanned. In both cases, sinusoidal interference fringes are obtained in the two-photon coincidence rates corresponding to the four possible detector combinations. This underlines that entanglement is invariant under analysis basis rotations. Error bars assume poissonian statistics.\label{Fig4}}
\end{center}
\end{figure}

Entanglement analysis is performed by recording interference fringes while scanning Bob's phase $\phi_{\rm i}$ for two fixed values of Alice's phase $\phi_{\rm s}=0$ and $\phi_{\rm s}=-\frac{\pi}{2}$, respectively.
As shown in \figurename~\ref{Fig4}(a-b), sinusoidal fringes are obtained for all the four two-photon coincidence configurations. As expected, pairs of detectors measuring identical polarisation modes (SPD$_{{\rm s},V}$ \& SPD$_{{\rm i},V}$ and SPD$_{{\rm s},H}$ \& SPD$_{{\rm i},H}$), as well as pairs measuring opposite modes (SPD$_{{\rm s},V}$ \& SPD$_{{\rm i},H}$ and SPD$_{{\rm s},H}$ \& SPD$_{{\rm i},V}$) are always in phase. The fact that in both cases, sinusoidal interference fringes are obtained, demonstrates the invariance of entanglement under rotations of the analysis basis.
From fitting the data with sinusoidal curves we extract the following fringe visibilities:
For Alice at $\phi_{\rm s}=0$, we obtain $82.6 \pm 1.3\%$, $82.9 \pm 1.3\%$, $87.7 \pm 1.2\%$, and $84.2 \pm 1.3\%$ for the detector pairs SPD$_{{\rm s},V}$ \& SPD$_{{\rm i},V}$, SPD$_{{\rm s},H}$ \& SPD$_{{\rm i},H}$, SPD$_{{\rm s},V}$ \& SPD$_{{\rm i},H}$, and SPD$_{{\rm s},H}$ \& SPD$_{{\rm i},V}$, respectively.
At $\phi_{\rm s}= -\frac{\pi}{2}$, we obtain $82.8 \pm 1.4\%$, $84.4 \pm 1.6\%$, $90.5 \pm 1.5\%$, and $86.3 \pm 1.5\%$ for the detector pairs SPD$_{{\rm s},V}$ \& SPD$_{{\rm i},V}$, SPD$_{{\rm s},H}$ \& SPD$_{{\rm i},H}$, SPD$_{{\rm s},V}$ \& SPD$_{{\rm i},H}$, and SPD$_{{\rm s},H}$ \& SPD$_{{\rm i},V}$, respectively.

When Bob's phase settings are $\phi_{\rm i} = \frac{\pi}{4}$ and $\phi_{\rm i} = \frac{3\,\pi}{4}$, respectively, we can extract the necessary expectation values for the violation of the Bell inequalities via the $S$-parameter~\cite{Bell_EPR_1964,CHSH_1969,Nielsen_book_2011}.
For $|S|>2$, nonlocal correlations associated with the created state are generally proven. The difference to the optimal value of $2\,\sqrt{2}$ can be used as a measure of the quality of the entangled state.

From our raw data, we extract $S_{\rm raw} = 2.50 \pm 0.02$ which represents a violation of the Bell inequalities by more than 25 standard deviations.
Additionally, when subtracting noise originating only from the detectors' dark counts ($\sim$30 counts per 2 seconds), we obtain $S_{\rm net} = 2.75 \pm 0.02$, which is only 3.5 standard deviations away from the optimal value of $2\,\sqrt{2}$. This further underlines the quality of our source and showcases the results that could be achieved using state-of-the-art superconducting single photon detectors~\cite{Hadfield09}.

\subsection{Performance evaluation}

In order to characterise the performance of a photon pair source, two parameters are critical.

First, the achievable photon pair generation rate for the maximum reasonable pump power at which no material damage occurs and at which the experiment remains stable.
Thanks to their high efficiency, standard $\sim$4\,cm long type-0 PPLN/Ws pumped at a few 10\,mW usually generate a stable flux of $0.1 \rm\,\frac{photon\,\,pairs}{coherence\,\,time}$ (independent on the filtered spectral bandwidth), which becomes particularily interesting for experiments requiring ultra narrowband photon pairs~\cite{Halder_source_2008,Kaiser_source2}.
For experiments based on discrete variables, higher photon pair generation rates are usually not considered as the experimental results become severely compromised by multipair emission contributions.
Moreover, in our case with rather broadband photons, we are constrained by the detectors' maximum counting rates way before multipair contributions become an issue, and therefore the waveguide-coupled pump power in our experiments never exceeds $50\rm\,\mu W$.
Consequently, we conclude that the maximum achievable pair generation rate is generally not a concern for PPLN/W based photon pair sources.
The performance of our particular PPLN/W compares a little bit unfavourable to state-of-the-art PPLN/Ws~\cite{Kaiser_source2}.
We measure a downconversion efficiency of $\eta \approx 2.5 \cdot 10^{-6}\rm\,\frac{generated\,\,photon\,\,pairs}{waveguide-coupled\,\,pump\,\,photon}$ within the 40\,nm broad emission spectrum of the source.
In other words, the internal spectral brightness of our source is $\approx 1.96 \cdot 10^6$ generated photon pairs per second, per milliwatt of pump power and per GHz of photon pair spectral emission bandwidth.
Therefore, $0.1 \rm\,\frac{photon\,\,pairs}{coherence\,\,time}$ are generated at a waveguide-coupled pump power of $\sim$116\,mW. For our MgO doped PPLN/W, this power level is still reasonably low and can be coupled into the PPLN/W without compromising its performance and stability.

Another important criterion is photon pair propagation loss from the source to the fibres which arrive at the users. In our case we a measure single photon propagation loss of $\approx 6.5\rm\,dB$ from the output of the PPLN/W to Alice's and Bob's analysers (\textit{i.e.} in front of the EOMs), respectively. The main contribution comes from our rather lossy 1\,nm bandpass filters ($\approx 4\rm\,dB$) which could be reduced by more than 3\,dB with state-of-the-art dense wavelength division multiplexers (DWDM).
With such DWDMs, one could use our polarisation entangled pair source as a heralded single photon source with a heralding efficiency of $\approx 50\%$. This would be amongst the highest ones reported for photon pair generators based on guided-wave photonics~\cite{McMillan_heralded_fibre_2009,Krapick_heralded_ppln_2013,Lutfi_heralded_ppln_2015} and further underlines the versatility of our source.

\subsection{Chromatic dispersion measurements}

We now demonstrate how this setup can be further exploited for measuring chromatic dispersion in the PMFs used in the Sagnac loop.
As signal and idler photons are non-degenerate, their wave-packets propagate at different group velocities in all the (fiber) setup.
In the standard configuration of the experiment, the optical path lengths from the source to the detectors are identical for both contributions, \textit{i.e.} $|V \rangle_{\rm s} |V \rangle_{\rm i}$ and $|H \rangle_{\rm s} |H \rangle_{\rm i}$. Therefore, signal and idler photons arrive always with the same time delays at their respective detectors, no matter which of the two contributions is analysed.

If, however, the length of one fibre in the Sagnac loop is considerably increased, say \textit{e.g.} the right hand side fibre in which the contribution $|H \rangle_{\rm s,\circlearrowright} |H \rangle_{\rm i,\circlearrowright}$ propagates, then this contribution leads to an arrival time difference which is greater than the one related to $|V \rangle_{\rm s,\circlearrowleft} |V \rangle_{\rm i,\circlearrowleft}$. We now denote $\Delta t$ as the related additional time difference introduced between the photons in the $|H \rangle_{\rm s,\circlearrowright} |H \rangle_{\rm i,\circlearrowright}$ contribution.
Note that, in a previous study, the resulting broadening of the coincidence peaks has been used for evaluating chromatic dispersion in the employed fibres with a sensitivity related to the timing resolution of the single photon detectors $(\approx 100\rm\,ps)$~\cite{Brendel_dispersion_1998}.

Here, we choose a different approach that achieves a resolution far below the detector timing jitter.
For this, we exploit the fact that the visibility of the interference fringes shown in \figurename~\ref{Fig4} dramatically depends on the degree of indistinguishability between $|H \rangle_{\rm s,\circlearrowright} |H \rangle_{\rm i,\circlearrowright}$ and $|V \rangle_{\rm s,\circlearrowleft} |V \rangle_{\rm i,\circlearrowleft}$ contributions.
Note that both contributions actually do not have to be distinguished by the detectors, the mere possibility to do so is sufficient to reduce the fringe visibility~\cite{HOM_dip_1987}.

\begin{figure}
\begin{center}
\includegraphics[width=1\columnwidth]{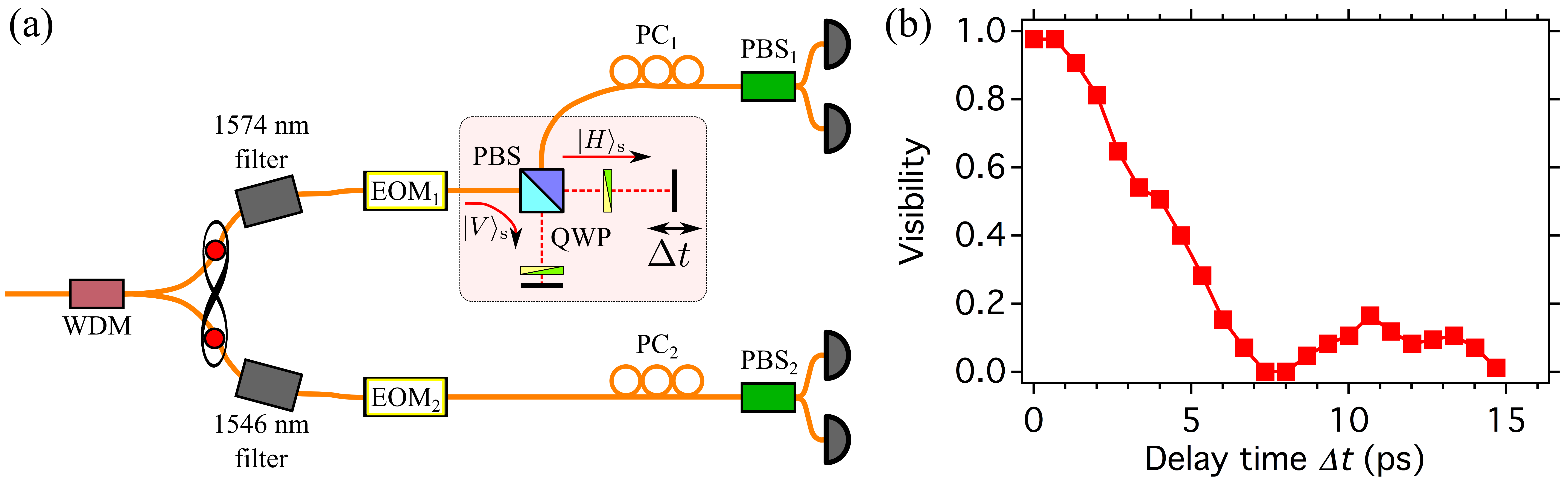}
\caption{(a) Experimental setup for referencing our experimental setup in the perspective of measuring chromatic dispersion via temporal delays. Compared to the setup in \figurename~\ref{Fig1}, we introduce a polarisation Michelson interferometer at Alice's analyser. This allows us, via a movable mirror, to introduce a controlled delay $\Delta t$ for the $|H \rangle_{\rm s}$ contribution through which a source of distinguishability is introduced. (b) Observed net interference fringe visibilities as a function of the delay time $\Delta t$. \label{Fig5_Reference}}
\end{center}
\end{figure}

Before measuring chromatic dispersion, we characterise first how the interference fringe visibility decreases as a function of the additional delay $\Delta t$. For this, we employ the exact same setup as used before for generating entangled photon pairs. However, as shown in \figurename~\ref{Fig5_Reference}(a), in Alice's path, a polarisation Michelson interferometer is introduced. This allows us, via a movable mirror, to introduce a controlled additional optical delay $\Delta t$ to the $|H \rangle_{\rm s}$ contribution.
The obtained (noise-subtracted) interference fringe visibilities as a function of $\Delta t$ are shown in \figurename~\ref{Fig5_Reference}(b).
For our photons, filtered down to $\sim$1\,nm spectral bandwidth, the fringe visibility drops down to 50\% at a delay of about 4\,ps which gives a temporal resolution comparable to state-of-the-art telecom measurement equipment~\cite{Gisin_CDcompare_1989}.
However, we mention that the resolution of our setup could be improved by 20 times, simply by removing the bandpass filters and therefore exploiting the full emission spectrum of the pair source.
Note that the data in \figurename~\ref{Fig5_Reference}(b) can neither be fitted by a Gaussian nor by a sinc-squared function. This indicates that our flat-top bandpass filters have rather smooth rising and falling edges and/or that the spectral phase of the photons is not flat within the transmitted spectrum.
Nevertheless, we have now obtained a precise \textit{ruler} which allows us to map measured visibilities to temporal delays.

\begin{figure}
\begin{center}
\includegraphics[width=1\columnwidth]{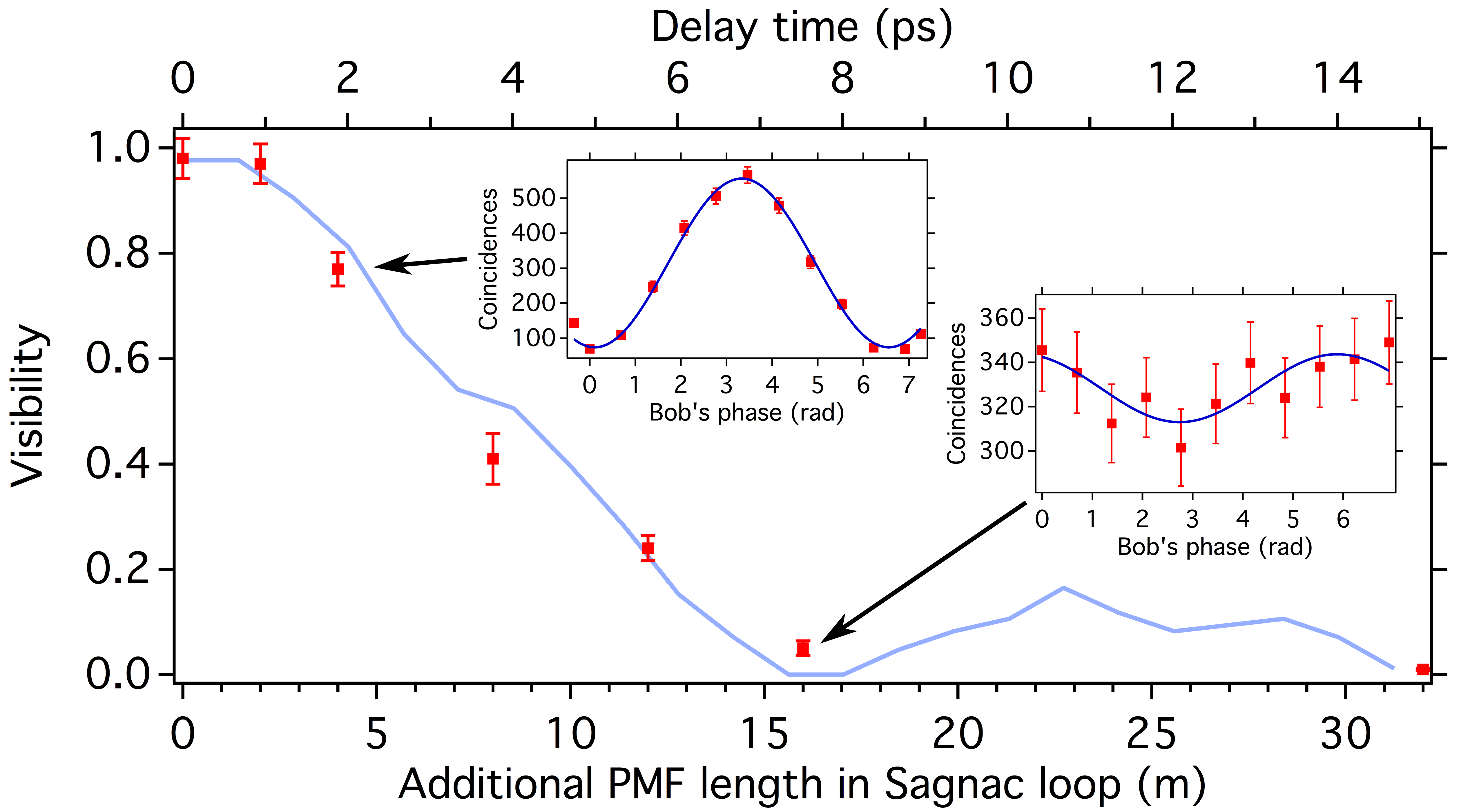}
\caption{Chromatic dispersion measurements via PMF length dependent interference fringe visibilities. Red dots represent measured visibilities as a function of the PMF length (bottom axis), and the light blue line represents the shape of the previously measured single photon temporal shape (top axis).
The scaling factor between the top and bottom axes is $\mathcal{R}=0.47\,\rm \frac{ps}{m}$, at which the sum of the error squares between both measurements is minimised. By normalising $\mathcal{R}$ with respect to the wavelength separation of the paired photons, we extract the chromatic dispersion coefficient $D = 16.79 \pm 2.14\rm\,\frac{ps}{nm \cdot km}$. The two insets show the recorded fringes obtained at $\Delta L = 4\rm\,m$ and $\Delta L = 16\rm\,m$.\label{Fig5}}
\end{center}
\end{figure}

We proceed now to measuring chromatic dispersion in the PMFs in the Sagnac loop. For this, set the delay in the polarisation Michelson interferometer to zero, and introduce additional PMF patch cords with length ranging from zero to 32\,m into the right-hand side of the loop.
We emphasise that the additional PMF delay does not have an influence on the delays of the pump laser photons as those have a coherence length of a few 100 metres. In other words, $|H \rangle_{\rm s,\circlearrowright} |H \rangle_{\rm i,\circlearrowright}$ and $|V \rangle_{\rm s,\circlearrowleft} |V \rangle_{\rm i,\circlearrowleft}$ contributions are generated from the same pump photon up to PMF lengths of $\sim$100\,m.
Experimental results are shown in \figurename~\ref{Fig5}. The red dots correspond to the data measured for different PMF length $\Delta L$ and the light blue line corresponds to the previously measured single photon temporal shape.
The value of chromatic dispersion in the PMF can now be extracted by inferring the scaling parameter $\mathcal{R}$ that maps a given additional PMF length $\Delta L$ to a temporal delay $\Delta t$:
\begin{equation}
\Delta t = \mathcal{R} \cdot \Delta L. \label{ScalingFactorEquation}
\end{equation}
By the method of the least-error-squares, we obtain $\mathcal{R} = 0.47 \pm 0.06\rm\,\frac{ps}{m}$ (the error corresponds to one standard deviation, \textit{i.e.} the offset at which the error square doubles compared to its minimal value).
The chromatic dispersion coefficient $D$ is then obtained by normalising $\mathcal{R}$ with respect to the wavelength difference between signal and idler photons $\Delta \lambda$:
\begin{equation}
D= 10^3 \cdot \mathcal{R} / \Delta \lambda.
\end{equation}
The prefactor of $10^3$ is introduced in order to comply with telecom standards where fibre lengths are usually measured in kilometres instead of metres.
In our case, the paired photons are separated by $\Delta \lambda = 28\rm\,nm$, and therefore we obtain $D=16.79 \pm 2.14\rm\,\frac{ps}{nm \cdot km}$ which is in very good agreement with previous measurements~\cite{Okamoto_Panda_Dispersion_1987}.

We mention that this method of measuring chromatic dispersion offers a great versatility in the sense that the sensitivity can be quickly adapted to the length and chromatic dispersion of the fibre under test.
Two procedures can be followed, which are generally very easy to implement.
On the one hand, higher (lower) sensitivity can be obtained by increasing (decreasing) the wavelength-separation of the paired photons, $\Delta \lambda$.
On the other hand, higher (lower) sensitivity could be achieved by reducing (increasing) $\tau$ using bandwidth-tunable filters.

Increased sensitivity would be required for measuring chromatic dispersion in short specialty fibres or fibres with nearly zero dispersion. On the other hand, reduced sensitivity allows measurements in rather long and/or highly dispersive fibres without the necessity to cut the fibre into shorter pieces.

Ultimately, by acquiring measurements at different $\Delta \lambda$, higher-order dispersive terms could also be inferred through the observation of a nonlinear dependence of $\Delta t$ as a function of $\Delta \lambda$.
For example, in a 0.3\,m long (nonlinear) optical fibre with ten times higher third order dispersion compared to a standard telecom single mode fibre at 1560\,nm, the observed fringe visibility at $\Delta \lambda \sim 300\rm\,nm$ would be 40\% instead of 42\% in a fibre with zero third order dispersion. Although such a small change in visibility cannot be observed with our current setup, this would be certainly feasible by employing high-speed superconducting single photon detectors and/or increasing integration times.

\section{Conclusion}

In summary, we have demonstrated a versatile fully guided-wave polarisation entangled photon-pair source based on a type-0 nonlinear waveguide mounted in a Sagnac loop.
A high degree of entanglement was observed, underlined by an $S_{\rm net}$-parameter of $2.75 \pm 0.02$, violating the Bell inequalities by more than 35 standard deviations.
The exclusive use of guided-wave optics, combined to the Sagnac-loop configuration, gives our source an excellent stability. Additionally, loss figures of merit are low and we have proposed a strategy for further optimization.
Although, here only energy-time entanglement has been revealed in a postselection-free fashion, we note that our source actually generates photon pairs which are hyperentangled in the polarisation and energy-time observables~\cite{Suo_hyper_Si_ring_2015,Kwiat_hyper_1997}. This makes our scheme interesting for high-dimensional quantum key distribution schemes, potentially combined with dense wavelength division multiplexing strategies towards realising high bit rate systems~\cite{Aktas_DWDM_2016,Lim2_Sagnac_2008,Autebert_DWDM_2016,Arahira_DWDM_2016}.

Furthermore, we have used our experimental setup for assessing chromatic dispersion in metre sized optical fibres via interference fringe visibility measurements.
In this perspective, our approach is very simple and flexible in the sense that measurement sensitivity can be easily adapted to the fibre under test. Moreover, higher order dispersion terms can be easily accessed through measurements at different wavelength separations. Combining this method with other interferometric quantum-enhanced chromatic dispersion measurement methods promises a great potential for high-precision fibre optical characterisations~\cite{Kaiser_Dispersion_2017}.

Finally, thanks to the exclusive use of guided-wave photonics, it should be possible to implement the full experimental setup into a standard 19-inch rack. In this case, our rather bulky pump laser system could be replaced by a compact laser diode, similarly as it has been done in a previous study~\cite{Stuart_FlexibleSagnac_2013}.

We therefore believe that our source is a good candidate for a broad range of applications in quantum information science.

\section*{Acknowledgement}

The authors acknowledge financial support from the Foundation Simone \& Cino Del Duca, the European Commission for the FP7-ITN PICQUE project (grant agreement No 608062), l'Agence Nationale de la Recherche (ANR) for the CONNEQT, SPOCQ and SITQOM projects (grants ANR-EMMA-002-01, ANR-14-CE32-0019, and  ANR-15-CE24-0005, respectively), and the iXCore Research Foundation.

\section*{References}

\end{document}